\documentclass[twocolumn,showpacs,preprintnumbers,amsmath,amssymb]{revtex4}
\usepackage{graphicx,color}   
\usepackage{amsmath}
\usepackage{dcolumn}
\usepackage{bm}

\begin{document}

\title{Characterizing the intrinsic correlations of scale-free networks}

\author{J. B. de Brito$^{1,2}$, C. I. N. Sampaio Filho$^1 \footnote{Correspondence to: cesar@fisica.ufc.br}$, A. A.
  Moreira$^1$, J. S.
  Andrade Jr.$^{1}$}

\affiliation{$^1$Departamento de F\'{i}sica, Universidade Federal do Cear\'a, 60451-970 Fortaleza, Cear\'a, Brazil \\
  $^2$Departamento de F\'{i}sica, Universidade Estadual do Piau\'i,
  64.002-150 Teresina, Piau\'i, Brazil}

\date{\today}

\begin{abstract}
  Very often, when studying topological or dynamical properties of
  random scale-free networks, it is tacitly assumed that degree-degree
  correlations are not present. However, simple constraints, such as
  the absence of multiple edges and self-loops, can give rise to
  intrinsic correlations in these structures. In the same way that
  Fermionic correlations in thermodynamic systems are relevant only in
  the limit of low temperature, the intrinsic correlations in
  scale-free networks are relevant only when the extreme values for
  the degrees grow faster than the square-root of the network
  size. In this situation, these correlations can significantly
  affect the dependence of the average degree of the nearest
  neighbors of a given vertice on this vertices's degree. Here, we
  introduce an analytical approach that is capable to predict the
  functional form of this property. Moreover, our results indicate
  that random scale-free networks models are not self-averaging, that
  is, the second moment of their degree distribution may vary orders
  of magnitude among different realizations. Finally, we argue that
  the intrinsic correlations investigated here may have profound
  impact on the critical properties of random scale-free networks.
\end{abstract}


\maketitle

\section{\label{sec:level1}Introduction} 

Very often, the network representation of a real system displays the
so called scale-free
property~\cite{barabasiRevModPhys2002,caldarelliPRL2002,boccaletti2006,barabasiScience2009}.
These networks typically exhibit a degree distribution in the form of
a power law, $P(k)\sim{k^{-\gamma}}$, with the most connected nodes
reaching a degree $k$ orders of magnitude higher than the average
degree $\langle{k}\rangle$. In fact, in most models of scale-free
networks the degree of the most connected nodes diverges at the
thermodynamic limit, $K_{max}\sim{N^\theta}$, with $N$ being the
number of nodes in the network and $\theta$ some positive exponent.
If $\gamma<3$ and with $K_{max}$ diverging, the second moment,
$\langle{k^2}\rangle$, also diverges, leading to some surprising
results~\cite{goltsevPRE2003}, such as the critical point of several
models in scale-free networks going to zero in the thermodynamic
limit~\cite{cohenPRL2000,cohenPRL2001,satorrasPRL2001,mendesPRE2002,satorrasPRE2002,cohenPRL2003,mendesRevModPhys2008}.

The exponent $\theta$ controlling the divergence of $K_{max}$ can also
affect significantly the structure of the
network~\cite{satorrasPRE2005}.  Simple extreme value
statistics~\cite{moreiraPRL2002,satorrasPRE2005,massimoPRE2014} shows
that, even if no other constraint is imposed, the most connected node
in the network should have a degree of the order of
$K_{max}\sim{N^{\theta_n}}$, with $\theta_n=1/(\gamma-1)$, which
represents the natural cutoff of scale-free
networks~\cite{satorrasPRE2005}. However, using the natural cutoff may
lead to unexpected properties in the network structure. Typically, if
the links are placed at random, one should expect the average number
of links between a pair of nodes do be given by
$n_{ij}=k_ik_j/N\langle{k}\rangle$~\cite{newmanPRL2002,newmanPRE2003}.
If the nodes $i$ and $j$ are extremely connected nodes,
one may expect their degree to be of the order of $K_{max}$. In this
case, for scale-free networks with $\gamma<3$, the maximum value for
$n_{ij}$ grows with the network size and should reach values above
one, that is, for these scale-free networks one should expect multiple
connections between the most connected nodes. However, it is usually
the case that network models do not account for multiple connections,
which would lead to the presence of intrinsic
correlations~\cite{satorrasPRE2005}. As a matter of fact, these
intrinsic correlations have been observed in different random network
models~\cite{satorrasPRE2005,gohEJP2006}. Moreover, since they can be
related to correlations of Fermionic systems due to the exclusion
principle, a practical analytical treatment becomes
possible~\cite{newmanPRE2003}. More specifically, in analogy
  with correlations of Fermionic systems that may become irrelevant at
  sufficiently high temperatures, the intrinsic correlations in
  networks may practically disappear if a more restrictive constraint
  is imposed to the extremely connected nodes,
  $K_{max}\sim{N^{\theta_s}}$, with $\theta_s\le{1/2}$. In this case,
  the maximum expected value for $n_{ij}$ does not diverge with $N$
  and multiple connections would not be present even if allowed. This
  so-called {\it structural cutoff}~\cite{satorrasPRE2005} represents
  a constraint that is strong enough to turn degree-degree
  correlations in the network negligible.

In this work we investigate the problem of intrinsic correlations in
random scale-free networks. This is performed by measuring the
dependence of the average connectivity of the nearest neighbors with
the degree of the node, $k_{nn}(k)$.  We confirm that for scale-free
networks the intrinsic correlations lead to a disassortative network,
with $k_{nn}$ being depreciated in the most connected nodes.  We
observe, however, that the particular form for $k_{nn}(k)$ varies
widely depending on the network instance, that is, two networks
of same size, generated with the same underlying degree distribution,
may display highly different dependencies for $k_{nn}(k)$. From our
results, we observe that the moments of the degree distribution,
$\langle{k}\rangle$ and $\langle{k^2}\rangle$, are determinant in the
shape of $k_{nn}(k)$. We then use an analytic approach, based on the
method of Fermionic correlations~\cite{newmanPRE2003} to obtain the
functional dependence of $k_{nn}(k)$ given the network size and degree
distribution. Since the moments of the degree distribution vary from
one realization of the network to another, we need to adjust the
minimum and maximum cutoffs of the degree distribution to each
realization. Using the properly adjusted degree distribution we
observe that our approach is capable of correctly predicting the
functional form of $k_{nn}(k)$. Our results show that random
scale-free network models are not self-averaging. More precisely,
relevant parameters of the network, such as the second moment of the
degree distribution, can vary orders of magnitude for different
realizations.

\section{\label{sec:level2}The intrinsic correlations}

To build our scale-free networks we use the configuration
model~\cite{bekessy1972,benderJCombThA1978,molloy1995,molloy1998}.  In
this model, we start by choosing the degrees of each of node from a
power-law distribution, namely, we let the degrees of the nodes to be
given by,
\begin{equation}
  k_i=\left\lfloor \left[ \left(K_{max}^{-\gamma+1}-K_{min}^{-\gamma+1}\right)r_i+K_{min}^{-\gamma-1}\right]^{\frac{1}{-\gamma+1}}\right\rfloor,
\label{kder}
\end{equation}
where $\lfloor\cdot\rfloor$ is the floor function and $r_i$ are random
numbers uniformly distributed between $0<r_i<1$. One should note that,
while the degrees $k_i$ are integers, the parameters $K_{max}$ and
$K_{min}$ may assume arbitrary values. After the degrees are
determined, the network is constructed by randomly selecting a pair of
nodes. The probability of selecting a particular node should be
proportional to its remaining degree. Once the pair is selected, a
connection should be placed between them and, if the connection is
allowed, their remaining degree is decreased by one. Note that we
include two important constraints, namely, there is no connection
between a node and itself, and there is no redundant connections, that
is, each pair of nodes will connect at most once. This later
constraint is similar to the exclusion principle of Fermionic
particles, being particularly important in scale-free networks with
$\gamma<3$ as it leads to the so-called intrinsic correlations.  The
imposed constraints may result in frustrated attempts to construct the
network because at some point, while constructing the network, one may
reach a situation in which all the nodes with remaining degree
are already connected. In this case, this attempt is rejected and the
whole process needs to be restarted by drawing a new set of degrees.
\begin{figure}[t]
\includegraphics[width=\columnwidth]{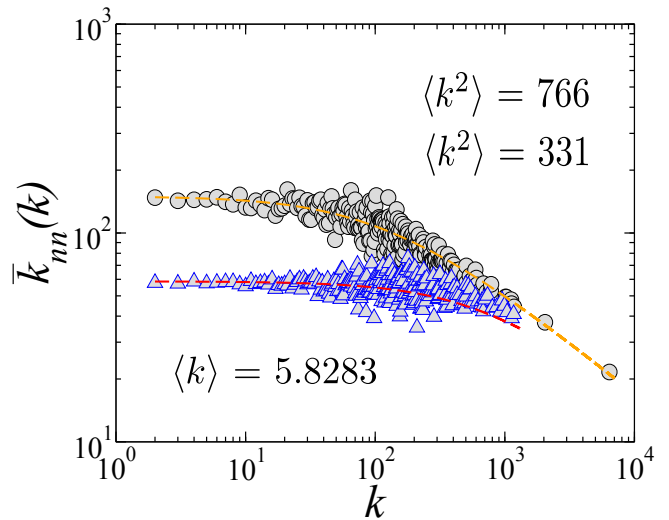}	
\caption{Average connectivity of the nearest neighbors of vertices
  with degree $k$, considering two values of $\langle{k^2}\rangle$ and the same value
  for the mean degree $\langle{k}\rangle$. The symbols represent simulations considering the
  configuration model~\cite{bekessy1972,benderJCombThA1978,molloy1995,molloy1998}. The dashed lines represent the
  results of the analytic approach described in the text. Here we considered $N = 10^{5}$ for the size
  of the network and $\gamma = 2.5$ for the degree distribution.}
\label{fig01}
\end{figure}

After the network is built, we proceed to calculate the average
nearest neighbor degree $k_{nn}(k)$. This is done by computing the
average degree of the nearest neighbors of each nodes, and then
averaging over nodes that possess the same degree. In Fig.~\ref{fig01}
we show the results obtained for two network realizations generated
with the same set of parameters, number of nodes $N=10^5$,
$\gamma=2.5$, and $K_{min}=2$.  We tune the parameter $K_{max+}=N-1$,
however, since frustrated attempts are discarded, the most connected
nodes of the network never reach this limit.  Curiously, these two
realizations of the model results in quite distinct forms for
$k_{nn}(k)$, although both present a clear disassortative trend, with
$k_{nn}$ decreasing for the most connected nodes. The difference
between these two realizations can be elucidated by noting that,
although having similar average degrees, they present very different
values for the second moment of the distribution, $\langle k^2
\rangle$.
\begin{figure}[t]
\includegraphics[width=\columnwidth]{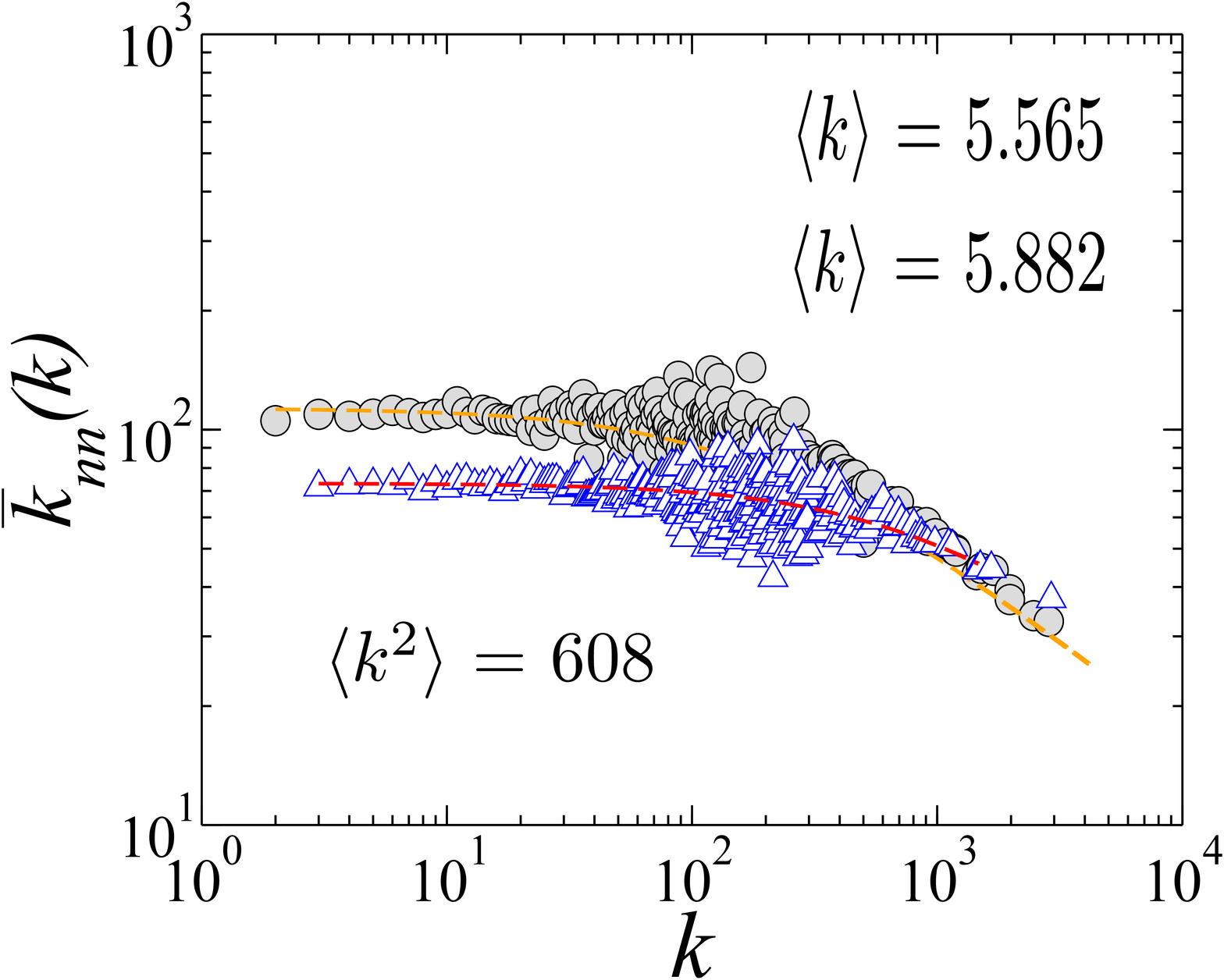}	
\caption{Average connectivity of the nearest neighbors of vertices
  with degree $k$, considering the same value for cumulant $\langle{k^2}\rangle$, and two
  different values for mean degree $\langle{k}\rangle$. The symbols represent simulations considering the
  configuration model~\cite{bekessy1972,benderJCombThA1978,molloy1995,molloy1998}. The dashed lines represent the
  results of the analytic approach described in the text. Here we
  considered $N = 10^{5}$ for the size of the network and $\gamma =
  2.5$ for the degree distribution.}
\label{fig02}
\end{figure}

\begin{figure}[t]
\includegraphics[width=\columnwidth]{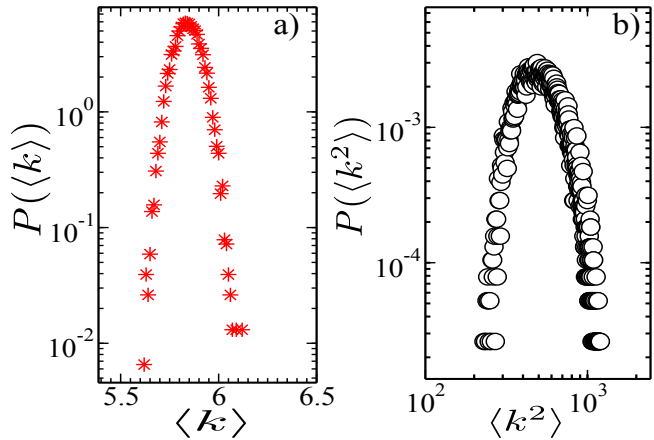}	
\caption{The probability distributions for the moments
  $\langle{k}\rangle$ and $\langle{k^2}\rangle$. Here we generate
  $10^6$ samples of networks from the configuration model, but do
  not allow for self and multiple connections.}
\label{fig03}
\end{figure}

In Fig.~\ref{fig02} we show another two realizations obtained with the
same conditions, but now we selected realizations with similar second
moments $\langle k^2 \rangle$, but distinct averages $\langle k
\rangle$. The results for $k_{nn}$ again do not match in this case,
although the difference is not as pronounced as in Fig.~\ref{fig01}.
The fact that the results vary from realization to realization
indicate that the network model is not self-averaging. In this way, to
understand the effect of the intrinsic correlation, we need more
information than just the network size and the value of the exponent
$\gamma$.  The absence of the self-averaging property can be observed
in Fig.~(\ref{fig03}), where the probability distributions for the
moments $\langle{k}\rangle$ and $\langle{k^2}\rangle$ are shown for a
set of $10^6$ samples of networks built from the configuration model
and considering the Fermionic conditions. It is important to notice
that the range of the $\langle{k^2}\rangle$ distribution covers two
orders of magnitude.

\section{\label{sec:level3}The self-consistent equations}

By considering that a network is uncorrelated results in an average
number of connections between a pair of nodes $i$ and $j$ given by
$n_{ij}=k_ik_j/N\langle{k}\rangle$. As mentioned before, this relation
cannot hold for scale-free networks with $\gamma<3$, unless some
restriction is imposed to the maximum degree of the nodes, $K_{max}$.
Our goal here is to be able to determine, given a network with
distribution $P(k)$ nodes of degree $k$, the dependence on $k$
of the average nearest neighbors $k_{nn}$.  The effect of intrinsic
correlations on random scale-free networks has been previously studied
in~\cite{newmanPRE2003}, where it was shown that the average number of
connections between a pair of nodes can be written as,
\begin{equation}
  n_{ij}=\frac{g(k_i)g(k_j)}{1+g(k_i)g(k_j)},
\end{equation}
where the form of the function $g(k)$ has to be determined by the
consistency of $k_i=\sum_{j}n_{ij}$. From that, one obtains the
following self-consistent set of equations for the values of $g(k)$:
\begin{equation}
  g(k)=k{\left(\sum_{k'}\frac{P(k)g(k')}{g(k')g(k)+1}\right)}^{-1}.
\label{selfcons}
\end{equation}
In order to obtain $g(k)$, we translate Eq.~(\ref{selfcons}) in
terms of an iterative process.
\begin{figure}[t]
\includegraphics[width=\columnwidth]{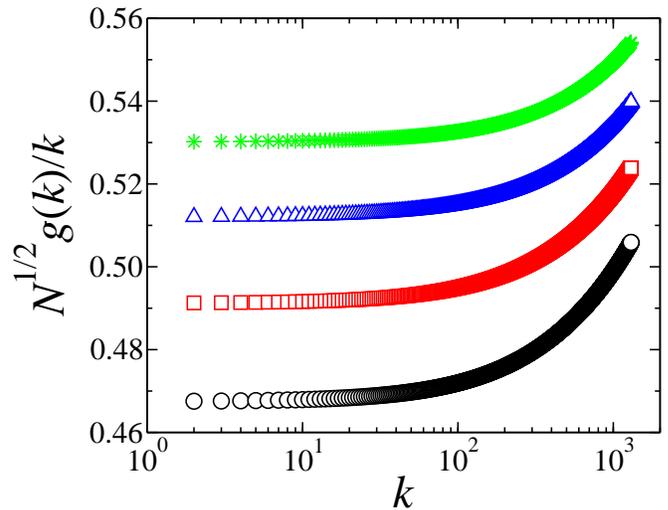}	
\caption{The $g(k)$ functions obtained from the self-consistent set of
  equations (\ref{selfcons}), as calculated for $K_{min}=2$ and $K_{max} =
  N^{1/(\gamma -1)}$. Here the following cases have been considered:
  $\gamma=2.5$ (circles), $\gamma=2.6$ (rectangles), $\gamma = 2.7$
  (triangles), and $\gamma = 2.8$ (stars).}
\label{fig04}
\end{figure}
We start with $g_0(k)=k/(N\langle{k}\rangle)^{1/2}$ as an tentative
guess for $g(k)$, and place $g_0(k)$ on the right side of
Eq.~(\ref{selfcons}) to obtain $g_1(k)$ on the left side. We then
proceed to obtain $g_2(k)$ from $g_1(k)$ and so on, until we a
self-consistent solution is obtained. In Fig.~\ref{fig04} we show the
form of $g(k)$ obtained through this process for different
expressions of the degree distribution $P(k)$. From $g(k)$, it is
possible to obtain the dependence of the average nearest-neighbor
degree as,
\begin{equation}
  k_{nn}(k)=\frac{g(k)}{k}\sum_{k'}P(k')\frac{g(k')k'}{g(k')(g(k)+1)}.
\end{equation}
However, as we observed from the results for the configuration
model, two important factors need to be taken into account. First,
some attempts to build the networks are rejected, therefore the
effective degree distribution $P(k)$ may be distorted by this
selection criteria. Second, since the model is not self-averaging, the
results for a given network may differ greatly from the average result
given the degree distribution used to generate the network. To
  obtain $k_{nn}(k)$, we need to determine the dependence we will
  introduce in the degree distribution $P(k)$ to match each network
  realization. The simplest approach is to set $P(k)$ exactly as the
number of nodes with degree $k$ that we obtained in a given
realization. Our simulations show that the obtained form for
$k_{nn}(k)$ follows closely the results observed with the
configuration model (data not shown).

Alternatively, instead of matching each value of $P(k)$ we tried to
reduce the number of parameters by matching only the first and second
moments of the degree distribution. To do that we observe that from
Eq.~(\ref{kder}) we obtain that, on average,
\begin{equation}
  P(k)=N\frac{k^{-\gamma+1}-(k+1)^{-\gamma+1}}{K_{min}^{-\gamma+1}-K_{max}^{-\gamma+1}}.
\end{equation}
However, considering that $K_{min}$ can be non-integer, the minimum
degree can be written as $k_{min}=\lfloor{K_{min}}\rfloor$, and
\begin{equation}
  P(k_{min})=N\frac{K_{min}^{-\gamma+1}-(k_{min}+1)^{-\gamma+1}}{K_{min}^{-\gamma+1}-K_{max}^{-\gamma+1}}.
\end{equation}
Therefore, by varying continuously the parameter $K_{min}$, we reduce
continuously the expected number of nodes with the minimum degree
$P(k_{min})$. Now, given the number of nodes $N$ and the exponent
$\gamma$, the two free parameters, $K_{min}$ and $K_{max}$, can be set
to match the values of the first and second moments obtained from a
particular random network realization. The results from the
configuration model presented in Figs.~(\ref{fig01}) and
(\ref{fig02}) are compared with the expected forms for the
average nearest-neighbor degree $k_{nn}(k)$ obtained with this
approach. As depicted, there is good agreement between the distinct
approaches.

\section{\label{sec:level4}Conclusion}

In this paper we have studied the intrinsic correlations of scale-free
networks considering a self-consistent set of equations $g(k)$, in
order to analyse the dependence of the average connectivity of nearest
neighbors with the degree of nodes. We have confirmed the
disassortative behavior of scale-free networks, and demonstrated that
the particular form for $k_{nn}(k)$ is strongly dependent on the two
first moments of the degree distribution. Moreover, using the properly
adjusted degree distributions, based on the minimum and maximum
cutoffs of the original degree distribution, we have observed
that out approach is capable to provide correct predictions for the
functional form of $k_{nn}(k)$. Finally, our results also
  indicate that random scale-free models are not self-averaging since
  these models are highly determinant parameters of the network, such
  as the second moment of the degree distribution, may vary orders of
  magnitude from one realization to the next. Therefore, it is likely
that our results have significant implications on the critical
behavior of models defined on complex networks
substrates~\cite{sooyeonPRE2007,motterPRL2007,mendesRevModPhys2008,mendesPRE2008,gleesonPRE2010}.

\textbf{Acknowledgments}

We thank the Brazilian agencies CNPq, CAPES and FUNCAP, the
National Institute of Science and Technology for Complex Systems
(INCT-SC Brazil.)

\bibliography{paper}

\end{document}